\documentstyle[12pt,version2,aps]{revtex}
\begin{document}
\renewcommand{\theequation}{\arabic{section}.\arabic{equation}}
\newcommand{\eqreset}{\setcounter{equation}{0}}
\setlength{\textheight}{20cm}
\vspace*{.9 in}
\begin{center}
{\large\bf
    STRONG-CORRELATIONS VERSUS U-CENTER PAIRING  \\
 and  FRACTIONAL  AHARONOV-BOHM EFFECT }\\
\vspace{.5 in}
{\sc F.V. Kusmartsev$^{*1,3}$, J.F. Weisz$^{**2}$ ,R. Kishore $^2$}

   and

 Minoru {\sc Takahashi $^3$}

 \vspace{.3 in}

{\it Department of Theoretical Physics}\\
{\it  University of Oulu}\\
{\it SF-90570 ~ Oulu, ~ Finland$^1$}\\

\vspace{.2 in}

{\it Instituto Nacional de Pesquisas Espaciaes-INPE}\\
{\it C. P. 515, 12201-970, S. J.  Campos, Sao Paulo, Brasil. $^2$}\\

\vspace{.2 in}

    {\it Institute for Solid State Physics,University of Tokyo}\\
        {\it Roppongi,Minato-ku,Tokyo 106 $^3$} \\

(received ~~~~~~~~~~~~~~~~~~
\end{center}

\vfill
\eject
\begin{abstract}

    The influence of the interaction between electrons on the
     Aharonov-Bohm effect is investigated in the
framework of the Hubbard model. The repulsion
between electrons associated with strong correlation is compared with the case
of attraction such as $U$-center pairing. We apply the Bethe
ansatz method
and  exact numerical diagonalization to the Hubbard hamiltonian. It is
shown
that the quasi half quantum flux periodicity occurs for any nonzero values of
$U
$ for two electrons. For large number of sites, or strong U, the quasi
periodicity becomes
an exact half quantum flux periodicity. However the character of the state
created on the ring is different in both cases.
In the case of $U$-center pairing the electrons are bound
in pairs located on the same site. For strong correlations(large positive $
U$) the electrons tend to be far from each other as possible.  We show
by numerical solution of Bethe ansatz equations that
for three electrons the flux periodicity of the ground state energy is equal to
1/3. The one third periodicity may occur even for small values of the ratio
$U/t$, for the very dilute system. It is shown  analytically using the
Bethe ansaz equations for $N$ electrons, that for dilute systems with arbitrary
value of the Hubbard repulsion  $U$,
the fractional Aharonov-Bohm effect occurs with period $f_T=1/N$ in units of
elementary
flux quantum. Such period occurs when the value $Nt/LU$ is small, where $L$
is the number of sites.  Parity effects dissapear in this fractional regime.
\end{abstract}

\vfill
\eject

\section{Introduction}\label{intro}

        The importance of strong correlations
in condensed matter physics became
obvious after the discovery
of the quantum Hall effect and high temperature superconductivity
\cite {Ande}. In fact the correlations are created by Coulomb forces
which are especially important
in low-dimensional systems \cite {Chak}-\cite{Ners}. In spite
of rapid progress in the understanding of strongly-correlated states such as
 the fractional quantum Hall effect\cite{Chak}, or the one-dimensional
Luttinger
liquid[3], the
microscopic detail of these correlations still needs to be furthur
developed\cite{KLN},\cite{Ners}.
 In order to shed light on their nature we investigate
using Bethe ansatz as well as by direct diagonalisation
an exactly solvable Hubbard model, describing a finite number of electrons
in magnetic field,
located on a ring.
 The filling factor is kept as a free parameter, by varying the number
 of sites. Lieb recently
pointed out, that  for small chains, the general ``Bethe
ansatz" solution, while correct, is too complicated for numerical
calculations \cite {Lieb1}.Examining this problem, we first investigate
the simplest case, when two electrons are located on the ring. In this case the
solution may be given analytically for certain cases.
We use the Aharonov-Bohm effect as a tool to study the pairing
   of the correlated state, and compare it with the $U$-center pairing.
In the limit
of large repulsion, when $U/t\rightarrow\infty$ Kusmartsev  \cite {Kus1}
found that the Aharonov-Bohm effect might be fractional. The Aharonov-Bohm
period is be changed from the conventional one, to $1/N$, where the $N$ is the
number of electrons on the ring.
Kusmartsev's result has since been confirmed
in other investigations \cite{Fye},\cite {Scho},\cite{Yu}. In the present work
we show that the fractional $1/N$
Aharonov-Bohm effect may occur for an arbitrary
 $N$  and for an arbitrary
(even very small)
ratio $U/t$ for the very dilute system, when filling factor drops to zero.

        The numerical calculations for a given number of sites are based on the
exact diagonalization of the two electron Hubbard hamiltonian. With this method
we can then analyze all the finite size effects, which may give
many instances of both level crossings and of
permanent degeneracies(as a function of $U$), as have been found by
Heilmann and Lieb \cite {Heil} in their studies of  the energy levels
of the benzene molecule having $6$ sites and $6$ electrons.

We obtain the same results
 from this direct diagonalization as from the Bethe-ansatz method.
\cite{Beth}-\cite{Lieb}.
 This contrasts with solutions of Bethe equations
obtained  in the thermodynamic limit  \cite {Lieb},  \cite{Lieb}\cite {Woyn2}
or calculations by
 Woynarovich and Eckle \cite {Woyn1}, who evaluated the
asymptotics of finite size effects on the ground state energy.
In fact, finite size objects such as rings may have
complicated   {\it  non-abelian}  dynamical symmetry groups  which were not
accounted for on the basis of the known invariance groups (spin,
pseudospin and  rotational symmetries of the ring) \cite {Lieb1}.
In fact  we found that finite scaling finite size effects
 were very important. There is a scaling behavior of ground state energy
which  does not depend on size $L$ or  on $U$ but which depends only on
$UL/N=const=\alpha$,
where $U$ is measured in units of $t$ and $L$ is measured in units
of lattice constant.  Such scaling occurs only at small values of $\alpha$.
Due to this scaling
symmetry  the fractional  Aharonov-Bohm effect may arise even for
small values of $U$, for very dilute electron systems.

\section { Model and Bethe equations}

To describe the fractional
Aharonov-Bohm effect for dilute systems we study the Hubbard hamiltonian
\begin{equation}
H = - t \sum_{<i,j>, \sigma} a^+_{i\sigma} a_{j\sigma}
     + U \sum^L_{i=1} n_{i+} n_{i-} \quad
\end{equation}
involving as parameters the electron hopping integral $t$ and
the on-site repulsive
Coulomb potential $U$ and the number of sites L.
The operator $a^+_{i\sigma} (a_{i\sigma})$  creates (destroys)
an electron
with spin projection $\sigma$ ($\sigma = +\, \rm{or}\, - $)
at a ring site $i$, and $n_{i\sigma}$
is the occupation number operator $a^+_{i\sigma} a_{i\sigma}$. The summations
in Eq. (1) extend over the ring sites $i$ or -- as indicated
by $<i,j>,\sigma$ -- over all distinct pairs of nearest-neighbor sites,
along the ring with the spin projection $\sigma$.
The effect of the transverse magnetic field is included via twisted boundary
conditions, with the Bethe ansatz substitution for the wave function.

     For the case of the magnetic field we will use the same form
of the wave function as has been proposed in the Refs \cite{Lieb},\cite{Suth}
\begin{equation}
\psi(x_1,\ldots,x_N)=\sum_P [Q,P]exp[i \sum^N_{j=1} k_{Pj} x_{Qj}]
\end{equation}

where $P=(P_1,\ldots,P_N)$ and $Q=(Q_1,\ldots,Q_N)$
are two permutations of $(1,2,..., N)$, and N is the number of electrons.

     The   coefficients  $[Q,P]$  as     well  as   $(k_1,...k_N)$   are
determined  from the Bethe equations which in a magnetic field  are
changed by the addition of the flux phase $2\pi f$ \cite{Suth}, \cite{Kus1}
\begin{equation}
e^{i(k_j L-2\pi f)}=\prod^M_{\beta=1}
 \biggl({{it \sin k_j -i\lambda_{\beta}-U/4}
\over{{it \sin k_j -i\lambda_{\beta}+U/4}}} \biggr)
\end{equation}
and
\begin{equation}
-\prod^N_{j=1} \biggl({{it \sin k_j -i\lambda_{\alpha}-U/4}
\over{{it \sin k_j -i\lambda_{\alpha}+U/4}}} \biggr)=
\prod^M_{\beta=1}\biggl({{i\lambda_{\alpha} -i\lambda_{\beta}+U/2}
\over{i\lambda_{\alpha}} -i\lambda_{\beta}-U/2} \biggr)
\end{equation}
The $f$ is the flux in units of elementary fundamental quantum flux $\phi_0$.
 The explicit form of Bethe equations in magnetic field is  \cite
{Lieb},\cite{Shas}, \cite{Kus1}, \cite{Fye},
\cite{Yu}
\begin{eqnarray}
Lk_j=2\pi I_j+2\pi f - \sum^M_{\beta=1} \theta(4(t\sin k_j -\lambda_{\beta})/U)
 \label{momenta}
\\
-\sum^N_{j=1} \theta(4(t\sin k_j -\lambda_{\beta})/U)=2\pi
J_{\beta}+\sum^M_{\lambda_{\alpha}=1}
\theta( 2(\lambda_\beta -\lambda_{\alpha})/U)\label{lambda}
\end{eqnarray}
where $\theta(x)=2 \arctan(x)$ and the quantum numbers $I_j $ and $J_{\beta}$,
which are associated with charge and spin degrees of freedom, respectively, are
either integers or half odd integers,
depending on the parities of the numbers of down and  up-spin electrons,
respectively:
\begin{equation}
I_j=\frac{M} {2} ~~~ {\rm (mod ~1) ~{and} }~~ J_{\beta}= \frac {N-M+1} {2}
{}~({\rm{mod} ~1).}
\end{equation}
The actual values (sets) of these numbers must be choosen to minimize the total
energy for the given
value of the flux $f$.

\section{ Half-flux periodicity for two electrons}

For the case of interest, for two electrons  on the ring
   the system of equations (3 and 4) simplify to two decoupled equations
\begin{equation}
L x = 2 \arctan{ \epsilon \over {\sin x}} + \pi n
\end{equation}
and
\begin{equation}
L y = 2\pi f + \pi m
\end{equation}
where $x= (k_1-k_2)/2$, $y=(k_1+k_2)/2 $
and $\epsilon= U/(4t\cos((2\pi f+\pi m)/L))$. The numbers $n,m$ may
have both positive
and negative values.

When $L=2$,   $(U>0)$, the
first of equations (5 and 6) may be solved immediately
and the solution is valid for the range of flux $|f|<1/2$, where
we must put $m=n=0$. For other
values of the flux the solution must be periodically
continued. The result is
\begin{equation}
k_{1,2}=\pm \rm{arccos}(-\epsilon/2+\sqrt{\epsilon^2/4+1})+\pi f
\end{equation}
which shows that with the increase of $U$ there is an increase of the
difference
between the $k$ vectors of the first and second electron. The increase of
this difference improves the quasi half quantum flux periodicity of the
Aharonov-Bohm
effect.  The ground state energy is described by simple expression:
\begin{equation}
E_{ground-2}= -2t (-\epsilon/2+\sqrt{\epsilon^2/4+1} )\cos(\pi f)
\end{equation}
\begin{equation}
j_{2}= -\partial E_{ground-2}/\partial f = -\frac {8\,\pi t^2\,
\sin (2 \,f\,\pi )}
    {\sqrt{{{\left(  {U^2} + 64\,{t^2}\,\cos (f\,\pi ) \right) \,} }}}
\end{equation}
where $f$ is limited to the region $|f|\leq 1/2$.
One sees that the value of the persistent current
at the fixed value of the flux  $f$ monotonically decreases as $U$ increases.
One sees that interactions change the current-flux dependence in a way
which is similar to that
of disorder or temperature, as shown in our recent paper
\cite {Weis}. That is, the interactions cause the
jumps in current-flux dependence,
or the cuspoidal points in the energy-flux curves disappear.
With stronger interactions these curves become gradually
smoother.
The energy dependence is a single-flux periodical function at any value
of $U$.
The reason for the single flux periodicity is that
the two site ring with two electrons is a very special case in that it is
half-full with two electrons.
In the limit $U\rightarrow\infty$
the current and the
energy vanishes, which coincides with the result obtained in Ref
\cite {Kus1}. That
is, for the half-filled
cases in that limit, the persistent current equals zero.

However, this single flux periodicity is broken, when we go away from
half-filling and immediately obtain quasi-half flux periodicity.
For example,  for the
Aharonov-Bohm effect on a ring having four sites (1/4 -filling)
the explicit formulas may also be written. In this case equation (5)
simplifies to the cubic equation
\begin{equation}
z^3+\epsilon z^2 -z -\epsilon/2=0
\end{equation}
where $z=\cos x$.  With the aid of the Cardano formula the solution can given
explicitly.  In the limit of
large value of $\epsilon$ it has the form:

\begin{eqnarray}
k_1(_2)= +(-)arccos(1/\sqrt{2}+1/(4\epsilon))+(2\pi f+\pi m)/4,~~ \rm {when}
{}~~\epsilon >>1
\label {mom-4sit}
\end{eqnarray}
with the flux energy dependence:
\begin{equation}
E_{ground-4}=-4 \cos( \frac {\pi f}{2}) (\frac {1} {\sqrt {2}}+
 \frac {1} {4\epsilon})
\label{4sit-en1}
\end{equation}
A nontrivial fact here is that there is another solution,
associated with singular values of $\lambda_\alpha$, which
for an arbitrary  value of $L$ has the form
\begin{equation}
E_{ground-L}=-4 \cos( \frac {2\pi (f - \frac  {1} {2}) } {L})
\cos(\frac {\pi} {L})
\label{Lsit-en2}
\end{equation}
 This formula coincides that describing the flux-energy dependence
of two noninteracting spinless fermions on a ring with $L$ sites.
For the problem under cosideration this dependence
also corresponds to a triplet state. This
means that on the two sites ring there
in a region of flux values near half-odd integer numbers in which
a very surprising
degeneracy between triplet and singlet states occurs. This means
that at this value of flux the matrix element for interaction
 vanishes for the singlet state.  For the singlet state there already occurs
a trapping of the flux quantum, which is shares between
two electrons.  The trapping of the quantum flux occurrs at each cuspoidal
point of the energy-flux curve. On the other hand
the trapping does not occur for the triplet state.
 This degenaracy, reflecting some hidden symmetry,
may be schematically expressed with the aid of the formula:
 {\it singlet + flux quantum == triplet}.

For the two sites ring discussed above, this solution
(\ref {Lsit-en2}) exactly equals  zero, giving a single flux periodicity
in the flux energy dependence for the half-filled ring.
Thus, for a non-half-filled ring the ground state flux-energy
dependence is determined by two solutions (\ref{4sit-en1}) and
(\ref {Lsit-en2}), which give  the needed quasi-half periodicity.

 From the equations (\ref  {mom-4sit})
one sees that the phase increases with $U\rightarrow\infty$ from
zero to $\pi/4$.
        The general result, which is valid for any value of  $L$, is as
follows:
With the increase of  $U$ the difference between $k_1$ and $k_2$ increases, and
the solution is given by the following formula
\begin{equation}
k_{1,2}=(\pm\theta+2\pi f +\pi n)/L
\end{equation}
where the function $\theta$ depends on $\epsilon$ and increases with $U$ and
$L$
. For the case $L=2$ the explicit dependence $\theta(U,f)$ is given
by equation (7).

The ground state energy is determined from  the formula
\begin{equation}
E=-2t \sum_{j=1}^N\cos k_j
\end{equation}
where the Beyers-Yang theorem \cite {Baye},\cite {Yang2} ( see also Refs
\cite{Suth}, \cite {Kus})
 is used to remove flux.
 From equations (9) and (10) one may conclude
that for the ground state energy the
momenta $k_1$ and $k_2$ must be single quantum flux periodical functions.
Because of the difference between
$k_1$ and $k_2$, due to the function $\theta$,
the ground state energy $E$ as a function of flux becomes a
quasi half flux periodic function.

With the increase
of $U$ or $L$ ( $L>2$) the shift $\theta$ increases; and consequently the
 half quantum flux
periodicity improves. This effect exists in two cases.
Good half flux quantum
periodicity appears for a small ring with a large U or for a
large ring with small $U$. Numerical results indicating these effects are
presented in Figs. 1 and  2.

For the electrons on the Hubbard ring the repulsive potential
$U$ causes the particles to locate on opposite sides of the ring. Because of
finite
 size effects(or alternatively the kinetic energy of electrons)
this localization
 is not complete. The criteria for strong $U$ to have good quasi half
quantum flux periodicity is then that $U$ is much larger that the kinetic
energy.
This is why the half flux periodicity improves both with
larger $L$ and larger $U$.

        It is interesting to compare the results with the Aharonov-Bohm effect
in
the case when there exists a pairing of the electrons induced by a negative
    $U$-potential (Pairing due to a negative $U$ center). In a one dimensional
system
with negative $U$ pairing one expects that the two electrons will tend
   to pair together on the same site. On the ring these electrons will also
have
    the tendency to move in pairs. The kinetic energy, due to the finite size,
will
 try to destroy the pairs. Therefore we again have in this case the approximate
half quantum flux periodicity. For the same reason as in the correlated
    state discussed above, the half quantum flux periodicity is improved with
the
    increase of $|U|$ and $L$. However the character of this state is different
   from the correlated one. For large negative $U$ one needs an activation
energy
(spectral gap) to destroy the localized
pair.

        For illustration we show the ground state energy dependence on the
   flux in two cases: constant value of $U$ with increase in the number of
sites
   ; and at constant value of $L$ with the decrease of the negative value of
$U$
   . We see(Fig. 3) that in both cases half quantum flux periodicity improves
as
    both $|U|$ and
$L$ increase.
The change in slope of the ground state energy as a function of flux will
correspond
to change in the direction of persistant current. In ring superconductors
 this current keeps flux quantized in units of half quantum flux, for which
   the negative $U$ case is a plausible model. The magnetization also behaves
similarly
to the current.

\section{ 1/3 flux periodicity for 3 electrons}

In contrast with the previos section, the three body problem does not allow
an explicit solution. As
discussed by Lieb\cite{Lieb1} the direct  numerical solution of Bethe
ansatz equations for small chains is a much difficult problem than the case of
the thermodynamic limit.
However, the problem for finite chains may be solved
if we give the Bethe ansatz equations in a
form convenient for  numerical iteration procedure.
The problem is to solve these equations
(\ref {momenta} ) and (\ref{lambda} ),  that is
to find numerically the values of
the variables $k_j$ and $\lambda_\alpha$ .
In order to find such solutions we  must represent the Bethe equations
in  a form convenient for iteration procedure, which is usually used
in numerical calculations.
 The first equation (\ref{momenta}) is already in the required form if we
divide both sides by
$L$. In the second equation
(\ref{lambda} ) we add to both sides the function
$ N\theta(4\lambda_{\beta}/U)$.
With the aid of these tricks one reduces the second equation
(\ref{lambda}) to the form:
\begin{equation}
\lambda_{\beta}=\frac{U} {4} \tan \big[\frac{
N \theta( \frac {4\lambda_{\beta}} {U})+
2\pi J_{\beta}+\sum^M_{\lambda_{\alpha=1}}
\theta( \frac { 2\lambda_\beta +2\lambda_{\alpha} } {U})-\sum^N_{j=1}
\theta(\frac {4t \sin k_j - 4\lambda_{\beta}}{U })} {2N} \big]
\label{lambda-res}
\end{equation}
 With the substitution
$\lambda_{\beta}= t_{\beta} U/4$ this equation
may be simplified and we arrive at a
couple of equations, convenient for the iteration procedure:
\begin{eqnarray}
k_j= \frac{2\pi I_j+2\pi f - \sum^M_{\beta=1} \theta(4t\sin k_j/U -t_{\beta})}
{L}
\label{momenta-res}
\end{eqnarray}
and
\begin{eqnarray}
t_{\beta}=\tan \big[\frac{ N \theta(t_{\beta})+
2\pi J_{\beta}+\sum^M_{\alpha=1}
\theta( (t_\beta -t_{\alpha})/2)+\sum^N_{j=1} \theta(4t\sin k_j
/U-t_{\beta})}{2N}\big] \label{iteration-res}
\end{eqnarray}
In what follows below we  use $U$ to represent the ratio $U/t$.
We have solved these equations iteratively, for the case of 3 or 4 electrons
on the ring, for different values of $F$,$U$ and $L$.
The convergence  depends on the value of the parameters $U$ and $L$ and
is fast if the value $U$ or $L/N$ are large.
The results may be classified as follows.
For small number of sites, and small value of $U$,
the ground state energy dependence
remains that of the free particle case.
The ground state energy dependence on flux
is a single flux periodical function. The ground state energy corresponds to
the case when two particles have down-spin  and particle  has an up-spin, or
alternitively, two particles
have up-spins and one particle down-spin. In that case the energy decreases
monotoneously
when flux increases from zero upto $f= 0.295167$ and then energy increases when
flux increases
upto 1/2. This behavior must be symmetrically reflected on the second half of
the
elementary flux unit and
then periodically continued for an arbitrary values of the flux. If
the value of $U$ becomes larger
than the critical value, which, for the ring of 4 sites is equal
to $U_c\sim 20$,
there appear new minima at integer values of the flux in the energy dependence.
In Fig.5 this dependence
is  calculated for  $U=50$. The shape of
that curve is very different from the free
fermion case. Each of these parabolic curves is associated with the state
characterized by a definite set of quantum
numbers $I_j$ and $J_\alpha$.
To show the validity of the Bethe ansatz equations
for the value $M>N/2$ we have solved these equations for two
cases, associated with the values of $M=2$ and $M=1$. For
these cases  the Bethe equations
have different forms although, physically, the states are equivalent
to each other.  In fact, the flux-energy dependency
for these cases are identical, although the quantum
numbers are distinct.
 Each state associated with the parabolic curve on Fig.5
 is represented
by a vertical column in the Table
\ref {tab}. One sees that for the transition of
one state to the other at the value $M=2$, the set of quantum numbers
is changed drastically.

\begin{table}
\caption{
  The  sets of quantum
numbers $I_j$ and $J_{\alpha} $ correponding to
the parabolic curve with the lowest energy and the value
of the flux at the minimum values are given in the vertical column.
 }
\begin{tabular}{ccccc}
\hline
 State's Number &1                                    & 2  &  3 & 4   \\
\hline
$M=2$                  $I_j $ &  (-1,0,1)  &  (-1,0,1)   &  (-2,-1,0) &
(-2,-1,0)\\
\hline
 $M=2$            $J_{\beta} $  &   (1,2)    &  (-1,0 )  &  (0,1) &  (1,2)  \\
\hline
$M=1$                  $I_j $ &  (-3/2,-1/2,1/2)  &   (-3/2,-1/2,1/2)    &
(-3/2,-1/2,1/2)  &  (-3/2,-1/2,1/2) \\
\hline
 $M=1$            $J_{\beta} $  &   (3/2)    &  (1/2)  &  (-1/2) &  (-3/2)  \\
\hline
Flux      $f_{min}$    & 0  &  1/3   &  2/3  &  1   \\
\hline
\end{tabular}
\label{tab}
\end{table}
With further increase
of $U$ the minima
 become more profound, transforming gradually to the curve consisting of
equidistant
parabolas, which is the 1/3 flux periodical function.
However one gets the same effect with the increase of the number of sites $L$.
For example, for the value $U=50$ of
Fig.6  the flux-energy dependence is shown for  the 6 site ring. With
the increase of the  sites number from  $L=4$ to $L=6$ the 1/3 flux
quasi-periodicity
has been gradually increased.  If we take a smaller value of $U$,
for example $U=8$,
for the same number of sites  $L=6$, two parabolic curves
dissapear from the ground state
energy curve, which   now consists of only  two parabolas (see, Fig.7).
For a larger number of sites, $L=12$ for example
,one has again four parabolas for the
ground state energy and 1/3-flux-periodical
dependence also appears
( Fig. 8). It is clear from these calculations  that the general tendency
of the appearence $1/3$ flux periodicity of the ground state energy
 is either with the increase of $U$,  at the fixed $L$ or with
the increase of the value of $L$, for fixed value of $U$.

The appearence of the $1/3$ flux periodicity
may arise also at small values of $U$ ,
provided that the value $L$ is large. This is illustrated
in Fig.9 , where the value
of $U$ is equal to 1 and the number of sites is $L=128$.
One can claim even more, that the shape of the ground state energy-flux
dependence does not depend on on the particular values of $U$ and $L$, but
depends on their
product. Fig.10 shows the flux-energy
dependence for the value $U=0.5$ and the value $L=256$ with  the same
product $U L$=128 as in the former case, given in the Fig.9. The
comparison these two Figures,
which are identical if the energy scale is neglected,
allows one to conclude that there
is a scaling symmetry,
whereby the product $UL$ is constant and the shape of the flux-energy
dependence
is not changed. Note that to have $1/3$ flux periodicity this product must have
a large value.
The results of calculations we made for the case of four electrons ( $N=4$)
is to a large extent
the same, but with the difference that $1/4$ flux periodicity appears.

The general conclusion can be drawn that
 the fractional Aharonov-Bohm effect appears when
the parameter $\alpha=tN/UL$ is small, but not exclusively in the case
when $t/U$ is small, as discussed
in previous work by Kusmartsev \cite{Kus} and by Yu and Fowler \cite{Yu}.

Encouraged by these numerical investigations of very small rings
we investigate the general case of arbitrary number of electrons
on the ring, when $\alpha$ is small.  These parameter values correspond
to realistic situations when $U/t$ has a some fixed value
but the system has a
very dilute density.

\section{ Fractional 1/N flux periodicity for N electrons}

Let us show that the fractional
Aharonov-Bohm effect is created for an arbitrary
number of electrons $N$ on the ring and arbitrary values of $U$.
 From equation(\ref{momenta-res}) one sees that  the numerator
of the right-hand side
cannot be large than $2  \pi N$. This holds
since for values of quantum numbers satisfy
$I_j\leq N/2$, the flux $f<1$ and $\sum^M_{j=1} \theta( x_j)\leq \pi M$.
If $2\pi N/L<<1$ one
has values of $k_j<<1$. Hence, on the right side of the
equation (\ref{momenta-res}) the
value $4 \sin k_j/U\sim 4 k_j/U\sim N/UL=\alpha$, is a small parameter.
Therefore in zeroth approximation, for small $\alpha$,
the expression $4 \sin k_j/U$
may be neglected, and one gets an expression for
$k_j$ in the form:
\begin{equation}
k_j= \frac{2\pi I_j+2\pi f + \sum^M_{\beta=1} \theta(t_{\beta})} {L}
\label{momenta-lim}
\end{equation}
In an analogous way, from equation(\ref{iteration-res}) one gets an equation,
which does
not depend on $k_j$:
\begin{eqnarray}
N\theta(t_{\beta})=2\pi J_{\beta}+\sum^M_{{\alpha}=1}
\theta( (t_\beta -t_{\alpha})/2),
\label{spin-lim}
\end{eqnarray}
This coincides with the equation obtained by Yu and Fowler \cite {Yu},
 in the limit of large
$U/t$. From the equation  (\ref {spin-lim}) we calculate the sum needed for
the right hand side of equation (\ref{momenta-lim}). After the subsitution into
that equation one gets
\begin{equation}
k_j= \frac{2\pi} {L} ( I_j+ f + \sum^M_{\beta=1} J_{\beta})
\end {equation}
This  expression
was  first obtained in the limit $U/t\rightarrow\infty$  by Kusmartsev  \cite
{Kus1},
and then rederived by Yu and Fowler \cite{Yu}.

In our new derivation  we have not made any assumption about the value of  $U$.
The ground state energy is here $1/N$ flux periodical function, where the
ground state energy-flux
dependence consists of parabolic curves with minima at the flux value
$f_{min}=p/N$,
where $p$ is the number of parabolic curves.
\begin{equation}
E_{ground}=-D \cos \big[ \frac{2 \pi} {L} (f-\frac{p} {N}) \big],
\label{gsenergy}
\end{equation}
where flux  value $f$ is changed in the region $ (2p-1)/2N<f<(2p+1)/N$ and
$D=2 \sin (\pi N/L)/ \sin (\pi/L)$.
Also from the above derivation it is clear that the parameter of our expansion
is equal to
\begin{equation}
\alpha= \frac {N t} { L U} =\rho t/U
\label {parameter}
\end{equation}
where $\rho=N/L$ is
the filling factor.
Summarizing, for any fixed value of the ratio $t/U$,
there exists a dilute  density limit  associated with  $\alpha<<1$. In
this dilute limit
the conventional Aharonov-Bohm effect dissapears, and the fractional effect
takes over.
Let us note that for the  almost completely polarized system,  for example,
when number of up-spins is much
larger than the number of down-spins $M/N<<1$
the equation (  {\ref {spin-lim}) may be also solved analytically. In that case
one assumes that
the value of $t_{\beta}$  in the equation (\ref {spin-lim}) is small. This
gives  that
\begin{equation}
t_{\beta}=\frac {\pi} {N} J_{\beta}
\end {equation}
This looks like a
spectrum of free spinless fermions on the chain with $2N$ sites.

With the aid of the parameter defined above, one may find the first correction.
Making an  expansion in powers of the parameter $\alpha$,
and using the second Bethe equation (\ref {lambda}),  we get
the form:
\begin{eqnarray}
N\theta(t_{\beta}) -\frac{8} {U} \frac {1} {(1+t^2_\beta)} \sum^N_{j=1}
 \sin(k_j)=2\pi J_{\beta}+\sum^M_{\lambda_{\alpha}=1}
\theta( (t_\beta -t_{\alpha})/2),
\label{spin-lim-exp}
\end{eqnarray}
which may be reduced to the equation:
\begin{eqnarray}
N\theta(t_{\beta} -\frac{4} {NU} \frac {1} {(1+t^2 _\beta)} \sum^N_{j=1}
 \sin(k_j))=2\pi J_{\beta}+\sum^M_{\lambda_{\alpha}=1}
\theta( (t_\beta -t_{\alpha})/2),
\label{spin-lim-exp1}
\end{eqnarray}
With the substitution $x_\beta=t_\beta- \frac{4} {NU}\sum^N_{j=1}
 \sin(k_j)$ this equation is reduced to eq. (\ref {spin-lim}),
with unknown variables $x_\beta$. The  equation derived for the
variables $x_\alpha$ is independent of the  flux $f$  and the value of $U$.
It is just the equation
for an isotropic Heisenberg antiferromagnet on the ring having $2N$ sites
and $M$ down spins. The solution for $x_\alpha$ is independent
of the flux $f$ or the value $U$. However, the variable $t_\beta$,
expressed via $x_\alpha$ with the aid of the formula:
\begin{equation}
 t_\beta=x_\beta+ \frac{4} {NU}\sum^N_{j=1}
 \sin(k_j),
\label {sol-t}
\end{equation}
 does depend on both parameters : $U$ or $\alpha$ and $f$.
The first correction
to $t_\beta$ does not depend
on the index $\beta$. The  dependence of $t_\beta$ on the flux $f$
comes about through its explicit dependence on the momenta $k_j$
via the second term of the right hand
side of the eq. (\ref {sol-t}). Substituting
the equation for variables $t_\beta$ into
 the equation (\ref{momenta-lim}) for the momenta $k_j$ and making an
expansion using the parameter $\alpha$, we get the form:
\begin{equation}
Lk_j=  2\pi I_j+2\pi f +\sum^M_{\beta=1} \theta(x_{\beta}) +
        \frac {8   \sum^N_{l=1}(1-N\delta_{lj}) \sin k_l } {NU}
\sum^M_{\beta=1} \frac  {1} {1+ t^2_{\beta}}
\label{momenta-1order}
\end{equation}
For small parameter $\alpha$ this system of linear equations
may be solved, with the result:
\begin{equation}
k_j= \frac {2\pi I_j} {L ( 1+ \frac {8B}{UL})}
    +\frac {2\pi f} {L} +  \frac {2\pi } {NL}
      \sum^M_{\beta=1} J_\beta
 +\frac {2\pi } {L} \frac {8B} {(UL+8B)} \sum^N_{l=1} I_l
                               \label{momenta-2order}
\end{equation}
where $B=\sum^M_{\beta=1} \frac  {1}{1+ x^2_{\beta}}$ is a real number.
Taking this solution  into account, the formula for the ground state energy
takes on the form:
\begin{equation}
E_{ground}=- \tilde{D} \cos( \frac {2\pi} {L} \left(f- \frac {p} {N} +
\frac {8B} {(UL+8B)} \sum^N_{l=1} I_l \right))
\label {en-corr}
\end{equation}
where $p=- \sum^M_{\beta=1} J_\beta$ ,
$\tilde {D}=2 \sin (\pi N/ \tilde {L})/ \sin (\pi/\tilde {L})$ and
$\tilde{L}=L ( 1+ \frac {8B}{UL})$ .

Here the values of quantum numbers $x_{\beta}$ do not depend
on the magnetic field. Precisely speaking, they
do not change their values when the flux changes within a single parabola.
The ground state energy will be associated
with a new set of the quantum numbers $J_\beta$
This means that with the first correction taken into account,
these parabolic curves, which
the ground state energy consists of, change their position mostly
along the vertical axis but also slightly
along the horizontal axis, but do not change in form.
Thus, it is in this case, the quasi $1/N$ periodicity
is preserved.

      To conclude,  our investigation
sheds light on the case in which the ring contains
many electrons in the
limit of very dilute electron density.
In the correlated state an effective phase shift appears
between the momenta of the different electrons, a shift which is associated
with the repulsive interaction. Because of the shift the periodicity of the
Aharonov-Bohm flux may have a fractional value.
In contrast this effect is not expected to occur for the negative
$U$ center model, where
at best one will have only the half quantum flux
periodicity.

\section{ Low versus high density limit}

There is, however, a correspondence between the states associated with the
positive and negative $U$ values.
 The dilute density limit
of electrons on  the ring described by the Hubbard Hamiltonian with
positive values of $U$  correspond
to the high electron density limit, described by the
same Hubbard model with negative values of $U$, with the aid of the
transformation
\begin{equation}
c_{i,+} \rightarrow a_{i,+}
\end{equation}
and
\begin{equation}
c_{i,-}\rightarrow a+_{i,-}
\end{equation}
the Hamiltonian (1) transforms into
\begin{equation}
H=-t \sum_i ( a^+_{i,+} a_{i+1,+} - a^+_{i,-} a_{i+1,-}) +U \sum_i
n_{i,+}-U\sum n_{i,+} n_{i,-}
\label {tran-1}
\end{equation}
 Introducing  an axiliary field acting only on the spin-down
electrons with a flux through the
ring of $\Phi_{-} =\pi L $ the Hamiltonian (\ref {tran-1}) may
be transformed to the form:
\begin{equation}
H=-t \sum_{i,\sigma}  a^+_{i,+} a_{i+1,+}  +U \sum_i n_{i,+}-U\sum n_{i,+}
n_{i,-}
\label {tran-2},
\end{equation}
This is a negative $U$-center model in a magnetic field of
strength $U$, creating the
Zeeman term. Note that with this transformation
the number of spin-up particles is not changed
$N_{+,new}= N_{+}$,
but the number of spin-down particles is
equal to $N_{-,new}=L-N_{-}$.
In other words, spin-down particles are
equivalent to  holes in the original
Hamiltonian.
Thus the spin-up particles are moving in a field
of flux $ \Phi_+=f$ and spin-down particles are moving
in a field of flux
$\Phi_-=f+L\pi$.
 One sees also that the system with odd and even number of sites will
have different  energy-flux dependence. One of them will be
 transformed into the other one
with a shift of a half flux quantum.Therefore, there is a parity effect here.
 Let us now discuss the ring with an even number of sites. We  have shown
that  electrons on the ring,  described by the Hubbard
Hamiltonian with positive
values  of $U$  in low density limit, behave in the same way as
the high density electron case, described
by the Hubbard Hamiltonian with the negative values of $U$,
provided that the system is highly
polarized.

This comparison shows that the Zeeman energy term:
\begin{equation}
U=-\mu_B H \sum^L_{i=1} (n_{i+}-n_{i-})
\end{equation}
where $H$ is the magnetic field associated with the flux quantum threading
the loop, cannot be simply dropped, and in some cases may give
very interesting new physics, as, for example, fractional
Aharonov-Bohm periodicity for the model of negative $U$-centers.
In the latter case, one again has the fractional $1/N$ flux periodicity
of the ground state energy
and the persitent current. This the number $N$ is equal to the sum
of the number of
spin-up particles and the number of holes. Physically it is not clear why
the fractional $1/N$ periodicity would occurr in that model.
The detail investigation of the influence of the Zeeman energy on the
Aharonov-Bohm effect we postpone for a forthcoming paper,
assuming for now that this energy
is small (i.e. the ring has a very large radius).

\section{            Parity effects on a Hubbard ring}

    For spinless fermions there is a difference in responses
 to a magnetic field for the cases of even and odd number of particles
 on a ring \cite{Kus}. This is the so-called parity effect. The effect
 is practically unchanged if there is an interaction between
 these spinless fermions. When the number of spinless fermions on the ring
 changes from odd to even, there is
 a statistical half-flux quantum which shifts the energy-flux
dependence  by exactly half of the fundamental flux quantum. Therefore,
 for small values of the flux and at odd number of spinless
fermions, the ring behaves as diamagnet. When there is an even number
of particles it behaves a paramagnet.  Kusmartsev obtained this result by
exact solution with the aid of Bethe -ansatz, in the model of
interacting spinless fermions
on the ring \cite{23} \cite {Kus}.  This was also independently qualitatively
discussed by Legett  for general case ( called as Legett conjucture) \cite {25}
 and   was proven
by Daniel Loss \cite{Loss}, with the aid of bosonisation
method in the framework  of
the same model \cite {23},\cite {Kus} but for arbitrary coupling.
 However taking spins into account, the situation
 is drastically changed.

    Taking spins into account for noninteracting electrons gives
 the diamagnetic response only when there is $N=4n+2$ particles on
 the ring, where $n$ is an arbitrary integer. For all other cases
the response has a paramagnetic character. With finite temperature
and disorder there
 occurs a double parity effect,  in which,
for $N=4n+1$, and $N=4n$, the response is
paramagnetic; and for $N=4n+1$ and $N=4n+2$ the response is diamagnetic
\cite {Weis}.

    With the inclusion of the Hubbard interaction there appears an
 additional phase shift due to scattering of a given particle on the
 other particles, via two-particles iteractions. Each scattering event
 gives a phase shift $\theta(x)$ in the Bethe equations. For the case
 of spinless fermions,
 the parity effect is conserved, in spite of the appearance of
the new phases \cite{Kus}.

    However, with the Hubbard interaction, one has a different picture.
 The analytical solution  shows that the phase shift $\theta(x)$
 creates a quasi-half flux periodicity, which  improves when the parameter
 $\alpha$ becomes smaller and smaller. The interaction creates an
 additional to the statistical flux which appears between the two electrons.

    It is interesting that in the limit $U \rightarrow \infty$
 this phase shift is exactly equal to half-flux quantum. Therefore
 if the flux of the external magnetic field is equal to a half-flux quantum,
 then, with that additional statistical half-flux quantum , the
 total flux is equal to the unit of fundamental flux quantum.

    In comparison with the case of noninteracting
 electrons, where the periodicity is in units of flux quantum,
here we already have a periodicity  at half a flux
 quantum. For the three electron case the phase shift is different.
 The additional statistical flux arises on
counting one permutation with each of other electrons.
  The value of that phase can
  be estimated in the limit of $\alpha\rightarrow 0$ and equals
\begin {equation}
2\pi f_{stat} = \sum_{\beta=1}^M
\theta(t_{\beta})= \frac{2\pi}{N} \sum_{\alpha} J_{\alpha}
\end {equation}
where we get a fraction $f_{stat}=\frac{1}{N}$. In this case
one may think that this flux is attached to each electron, that
 is all $N$ electron  share 1 unit of the flux quantum. Putting
 a new electron on the ring creates a new system, where
 $N+1$ electrons will now share a unit of quantum flux. In
 this system, the response has a purely
 diamagnetic character, for any number of electrons.

 In general terms, the appearance of
 the parity effect and conventional Aharonov-Bohm effect
 may be described as follows:

    At small values of $U$, or more precisely speaking
 large $\alpha$, we have the conventional parity effect for free
 electrons (see, Table \ref {tab2}). With an increase of the
 interaction there exists the critical value of $U=U_{cr1}$ or
 $\alpha=\alpha_{cr1}$ where, for values  $U>U_{cr1}$
or  $ (\alpha<\alpha_{cr1})$ the parity effect looks similar to
 the parity  effect for spinless electrons.
 That is, for even number of electrons the magnetic response has
 a diamagnetic character and for an odd number of electrons
the response is paramagnetic. Note that for spinless fermions the
 response is diamagnetic for an odd number of electrons.
 With further increase of the coupling constant $U$ there is
 a second critical value,
at $U=U_2 ~~ (\alpha=\alpha_2)$, with the new type of
parity effect. Thus for $U>U_2 ~~ (\alpha<\alpha_2)$
the paramagnetic response occurs only for $N=4n+1$ electrons.
Finally, for $U>U_3$, so that $U_3>U_2$, the parity effect
disappears and the ring behave as diamagnet for any number of electrons.
The value of $U_3$ depends on the electron density. As
discussed above, the parameter $\alpha$ is what matters.
 From our investigation we may conclude that the critical value
for the disappearence of the parity effect equals $\alpha_3\sim0.02$
With the disappearence of the parity effect the quasi-$1/N-$ fractional
Aharonov-Bohm periodicity will appear. The classification of the different
parity regimes
are shown in the Table \ref {tab2}.

\begin{table}
\caption{
  The  table shows the classification
of different regimes of the parity effect,
with change of the Hubbard interaction. The notations {\it dia-} means that
the ring behaves as diamagnet, the notations {\it para-} means
that the ring behaves as a paramagnet. The number of electrons on the ring
is equal to $ 4n+2, 4n+1, 4n, 4n-1$ , respectively.
 }
\begin{tabular}{ccccc}
\hline
 Particle's Number &  $U<U_{c1}$ & $U_{c1}<U<U_{c2}$&   $U_{c2}<U<U_{c3}$ &
$U_{c3}<U$   \\
\hline
$4n+2$ &  {\it dia-} &  {\it dia-}  &  {\it dia-} & {\it dia-}\\
\hline
 $4n+1$ & {\it para-}   &  {\it para-} &{\it para-} &{\it dia-} \\
\hline
$4n$ &   {\it para-}  & {\it dia-}   & {\it dia-} &   {\it dia-} \\
\hline
 $4n-1 $  &   {\it para-}    &   {\it para-}  & {\it dia-} & {\it dia-} \\
\hline
\end{tabular}
\label{tab2}
\end{table}
The parity effect is preserved with  disorder or finite tempepature.
However the change of this classification
with the change of the temperature is nontrivial and will be discussed
in a forthcoming paper.

\section{Possible experiments}

     It is worth noting that there is a possible practical realization, where
the above model may be applicable; that is
the case of a ring consisting of quantum dots, placed in succession.
These structures may be built and investigated using modern technology
\cite {Deme}.
The single quantum dot will act as a potential well for the electrons. If the
radius
R of the quantum dot decreases, the charging energy  $e^2/R$  increases
and there occurs a case in which no more than two electrons with opposite spin
can be accommodated on a single dot. The system of quantum dots
is to be described with a Hubbard hamiltonian with $U=e^2/R$. Therefore in a
ring
consisting of quantum dots, with
high charging energy, one may observe the destruction of
the simple Aharonov-Bohm effect and the appearance of the fractional
period.

The effect of the fractional or $1/N$ periodicity is directly related
to the phenomena of Coulomb Blockade.  The novel feature of
both  these phenomena is due to
many body effects associated with the interacting current
carriers. That is, the motion or a change of state of a single electron,
changes the states of all other electrons.

 If we put the ring of quantum dots in a transverse magnetic field single
electrons will have the tendency to move along the ring. However hops
onto a dot already containing two electrons is not allowed, since this will
cost the energy equal to the charging energy. One may have incoherent
or independent hops. However, there is another possibility. This is a process
which will not cost the charging energy, in which all electrons on the ring
make
coherent hops (cotunneling). In other words this is the simultaneous motion
of all electrons on the ring. It is clear that if we move all
electrons together the charging energy is not important
($U$ may be arbitrarily large) and
the total change in the phase of the many-body wave function will be equal to
$ 2\pi fN$.

   From the guage invariance of the ground state of the ring
we conclude that the
equivalent state to $f=0$ is $N f=1$. Hence the period
of interacting electrons on the ring is $f=1/N$,
in agreement with the results obtained in this paper.

\section{Conclusions}

In the present work we have studied the effects of the electron-electron
correlations  on the Aharonov-Bohm effect in a quantum ring.
We have shown the correlations result in a fractional
Aharonov-Bohm effect, which appears
when the parameter $\alpha =Nt/LU$ is small. This case may
occurr when $U/t$ is large
or in low density limit when the filling  $N/L$ is small. The
 conclusion that the low density limit
of the Hubbard model
is equivalent to a strong coupling limit $U/t>>1$ coincides with one
obtained  by Shulz \cite {Shul},
a work which describes Luttinger liquid properties in the framework
of the bosonisation approach.
Using the  Aharonov-Bohm effect we proved this theorem
far beyond the low-frequency limit of that
theory.

We have found also a very intersting scaling symmetry, hold for the
low density limit, namely, that the shape of the ground and excited
flux energy dependencies depends only on the parameter $\alpha=\frac {tN}{UL}$.
In other words when $\alpha<<1$ the flux-energy dependencies obtained
for the different values $U$ and $L$, provided that the
parameter $\alpha=const$ is fixed, can be transformed one into another
by a scaling transformation of the energy scale. This confirms
the Lieb suggestion \cite {Lieb1} that the Hubbard Hamiltonian
describing a system of the finite size has some very nontrivial
internal symmetries.

 The density of electrons may be well controlled in many experimental
situations, for example, by doping, or by applying the gate voltage
to change of the position of the chemical potential
as it is used in quantum wells. Therefore the predicted
fractional Aharonov-Bohm effect is a good challenge
for experimentalists.

\vspace{3mm}
\begin{flushleft}
{\bf Acknowledgement}
\end{flushleft}
\vspace{2mm}

FVK and JFW  thank INPE and CNPq for financial support, during their stay at
INPE,S. j. Campos, Sao Paulo, Brasil.  The work by FVK  has been supported by
Faculty of Natural
Science of Oulu University and
by Ministery of Education, Science and Culture of Japan

\renewcommand{\theequation}{\Alph{section}.\arabic{equation}}

permanent address:

{\parskip=0pt
$*)$L.D. Landau Institute for Theoretical Physics \par
Moscow,117940, GSP-1, Kosygina 2, V-334, Russia \par }

\medskip
$**)$
INTEC(UNL-CONICET), Guemes 3450, 3000 Santa Fe, Rep. Argentina

\newpage
{\large \bf Figure Captions}\\

{\bf Fig.1 } \hspace{5mm}
 The ground state energy dependence on the flux of the external field
for the Hubbard ring at the fixed value of $U$ at the different values of $L$
Here $U=50t$ and $L$ is 3,4 or 6 sites. $\diamond$ is for $L=3$,  $+$ is for
$L=
   4$ and
$\sqcap$ is for $L=6$. Ground state energy at zero flux is subtracted off so
as to normalize the figures. Energies are in units of t and flux is in units of
quantum flux.

\bigskip

{\bf Fig.2 }The ground state energy versus external field flux
for the Hubbard ring at the fixed value of $L$ and different values of $U$
Here $L$ is 5 sites and $U=5t,20t,50t$. $\diamond$ is for $U=5t$, $+$ is for
$U=20t$ and $\sqcap$ is for $U=50t$.

\bigskip

{\bf Fig3.} The same as in Fig1., but for the case of $U-center$ pairing.
Energies are in units of t. Here $U=-10t$ and $L$ takes on values of
3,5 and 8 sites. $\diamond$ is for 3 sites, $+$ is for 5 sites and $\sqcap$
is for 8 sites.
Ground state energy at zero flux has been subtracted off to normalize the
curves. Flux is in units of the quantum flux.

\bigskip

{\bf Fig.4} The same as in Fig2., but for the case of $U-center$ pairing.
Here $L$ is kept fixed at 5 sites but $U=-1t,-5t or -20t$. $\diamond$ is for
$U=-t$, $+$ is for $U=-5t$ and $\sqcap$ is for $U=-20t$.

{\bf Fig.5}  The behavior of the ground state energy and the first excited
levels as a function of
flux for three electrons at the values $L=4$, $N=3$ and $U=50$ in the region
of flux within the single fundamental flux quantum. Two particles have
up-spins and one particle has down-spin.

{\bf Fig.6 } The same as in Fig.5 but at the value $L=6$.  At the cuspoidal
point there appears the flux absorption.

{\bf Fig.7 }  The same as in Fig.5 but at the value $L=6$ and $U=8$.

{\bf Fig.8 }  The same as in Fig.5 but at the value $L=12$ and $U=8$.

{\bf Fig.9 }  The same as in Fig.5 but at the value $L=128$ and $U=1$. The
energy is expressed in the units $t  10^3$ . The zero energy corresponds to
$-2.99 t$.

{\bf Fig.10}  The same as in Fig.5 but at the value $L=256$ and $U=0.5$. One
sees that the
shape of this dependence is scaled to the one shown in Fig.9. The energy is
expressed in the units $t  10^4$ . The zero energy corresponds to $-2.999 t  $.

\end{document}